\documentclass[9pt,twocolumn,twoside]{opticajnl}
\usepackage[]{graphicx}
\usepackage[]{amsmath}
\usepackage{jabbrv}

\journal{osajournal}
\setboolean{shortarticle}{false}

\title{Affine diffractive beam dividers}

\author[1]{F. Gori}
\author[2]{R. Mart\'{i}nez-Herrero}
\author[3]{O. Korotkova}
\author[2]{G. Piquero}
\author[4]{J. C. G. de Sande}
\author[1]{G. Schettini}
\author[5]{F. Frezza}
\author[1,*]{M. Santarsiero}

\affil[1]{Dipartimento di Ingegneria Industriale, Elettronica e Meccanica, Universit\`a Roma Tre, Via V. Volterra 62, 00146 Rome, Italy} 
\affil[2]{Departamento de \'{O}ptica, Universidad Complutense de Madrid, Ciudad Universitaria,  28040 Madrid, Spain}
\affil[3]{Department of Physics, University of Miami, 1320 Campo Sano Drive, 33146 Coral Gables FL, USA}
\affil[4]{ETSIS de Telecomunicaci\'{o}n, Universidad Polit\'{e}cnica de Madrid, Nikola Tesla s/n Campus Sur 28031 Madrid, Spain}
\affil[5]{Dipartimento di Ingegneria dell'Informazione, Elettronica e Telecomunicazioni, Universit\`a Sapienza, via Eudossiana 18, 00148 Rome, Italy}

\affil[*]{Corresponding author: msantarsiero@uniroma3.it
\newline
\newline
\bf © 2024 Optica Publishing Group. One print or electronic copy may be made for personal use only. Systematic reproduction and distribution, duplication of any material in this paper for a fee or for commercial purposes, or modifications of the content of this paper are prohibited.
\newline
\url{https://doi.org/10.1364/JOSAA.514290}
}

\begin{abstract}
 
Diffractive optical elements that divide an input beam into a set of replicas are used in many optical applications ranging from image processing to communications. Their design requires time-consuming optimization processes, which, for a given number of generated beams, are to be separately treated for one-dimensional and two-dimensional cases because the corresponding optimal efficiencies may be different. After generalizing their Fourier treatment, we prove that, once a particular divider has been designed, its transmission function can be used to generate numberless other dividers through affine transforms that preserve the efficiency of the original element without requiring any further optimization.
\end{abstract}
\begin{document}

\maketitle

%=====================
\section{Introduction}
\label{Introduction}
%======================

Diffractive optical elements (DOEs) use suitably designed diffractive structures aimed at altering the phase of an incident light wave in such a way as to obtain the desired irradiance distribution of the diffracted field, either in the near or in the far field. 
They can have optical functions that cannot be obtained, or could only be obtained with very complex techniques using traditional elements. 
For example, they can be used as beam shapers, beam dividers, pattern generators, light diffusers, and more. 
Comprehensive treatments of DOEs, their design, and applications can be found in~\cite{Loewen,Soifer}.

Subject of the present paper are multiple-beam dividers. 
{Such devices are used in many applications, such as material processing, interferometry, image processing, and optical communications~\cite{RD3}.}
Their study started from {Dammann} gratings and from deductions of upper bounds for their efficiency \cite{Damm,Jari,Wyro}.  For the simple case of the one-dimensional (1D) phase triplicator, a closed form expression for the optimum was found by means of the calculus of variations \cite{Gori1,Gori2}. The optimum phase duplicator with prescribed power ratio as well as the four-beam {divider} were also found in closed form \cite{Borghi,Borghi2}. The generalization of these results to {beam dividers of order $N$ (or $N$-plicators) of both 1D and 2D nature,  for several integer $N$ values,} was developed by Romero and Dickey in two masterful papers \cite{RD1,RD2}, whose content was later summarized in an article in which such a subject was seen in the general realm of beam shaping procedures \cite{RD3,Capasso}. Current important results of both theoretical and experimental nature have been obtained by Ignacio Moreno and his co-workers \cite{Moreno19,Moreno22,Moreno23}.
{It should be noted that beam dividers constitute, both conceptually and experimentally, the basic step in performing more sophisticated form of optical processing.}

In this paper, we revisit the design process of a diffractive beam divider. 
%First, attention will be called on the distinction between 1D and 2D cases because results obtained in the first case are not necessarily extendible to the second. 
{First, specific cases will be used to illustrate that the diffraction efficiency of 1D cases can be reduced for 2D cases and must be recomputed.}
Then, an affinity rule will be established, which allows us to design a lot of new dividers in a simple way on the basis of existing ones.

%=====================
\section{The model}
\label{model}
%======================

\begin{figure}[!ht]
\centering
{\includegraphics[width=6cm]{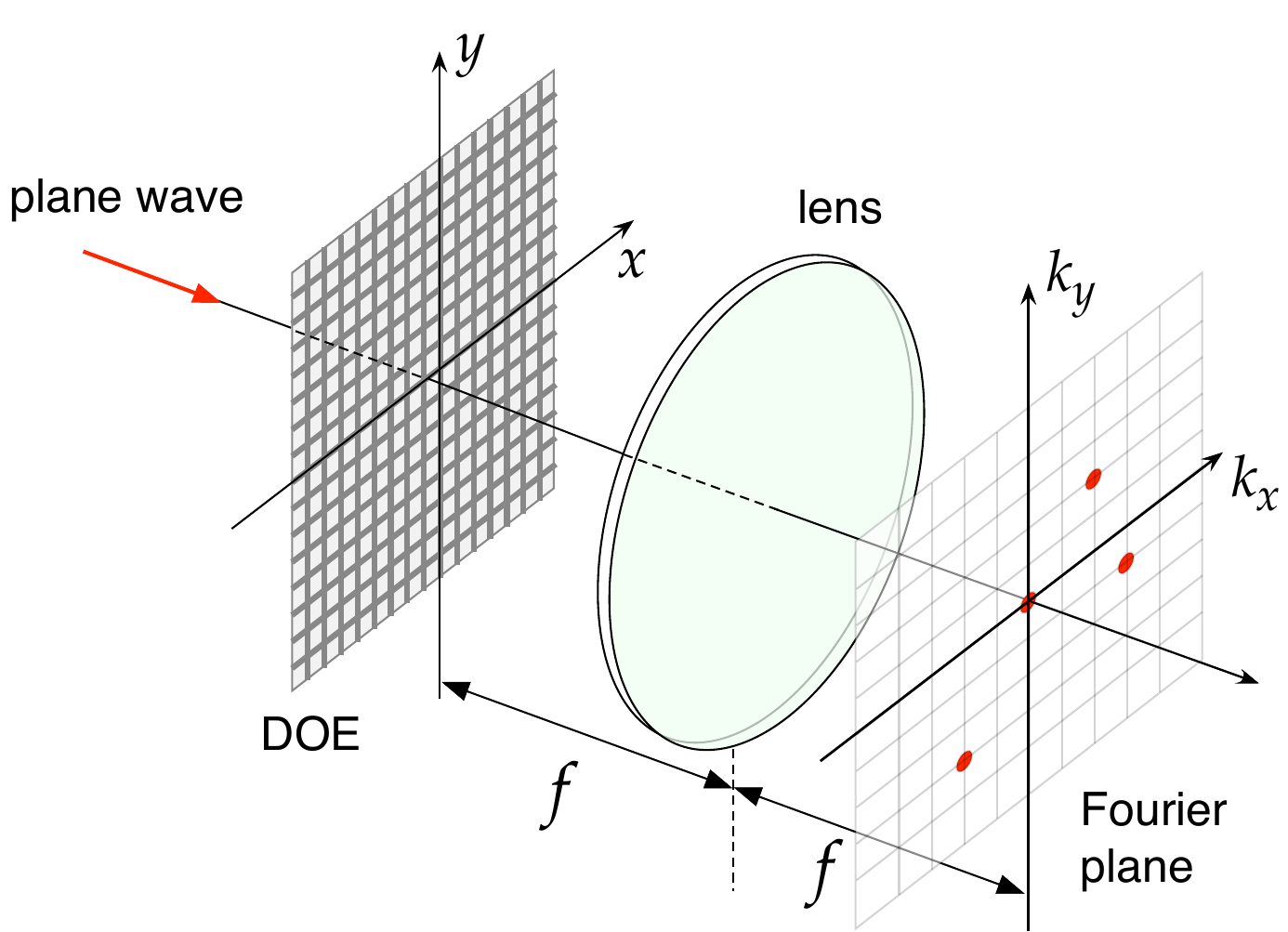}}
\caption{A beam divider produces N replicas of an incident plane wave, propagating along different directions.}
\label{figura1}
\end{figure}

We begin by stating the problem and specifying the model we are going to use to design the DOE.
Given a set of $N$ directions in space, we have to find a phase transparency that upon illumination by a plane wave produces $N$ plane waves proceeding along those directions with the maximum possible irradiances in prescribed mutual ratios. We shall limit ourselves to the case where all diffracted orders carry the same power, but the approach could be easily generalized. The scheme is sketched in Fig.~\ref{figura1}, where a $2f$ optical system is used to focus the diffracted plane waves onto the back focal plane of a lens.
The variables $x$ and $y$ denote the space coordinate across the plane of the DOE, while $k_x$ and $k_y$  give the $x-$ and $y-$components of a typical wave-vector.

{In order to find} the phase profile of the most efficient DOE we start from the field that would be produced by the interference of the $N$ diffraction orders across the plane $xy$. We assume that they produce a periodic field distribution $V(x,y)$ (for simplicity, without loss of generality, the period $P$ is supposed to be the same along the $x$ and $y$ axes), and this means that the $x$- and $y$-components of their wave-vectors are integer multiples of $K=2\pi/P$.

On expanding $V$ in a Fourier sum we have
\begin{equation}
\begin{array}{c}
\displaystyle   V(x,y)
= \sum_{(n, m)\;  \in B }   b_{nm} \, e^{i K (n x+ my)},
\end{array}
\label{new00}
\end{equation}
where $B$ is the set of $N$ pairs of indices ($n,m$) specifying the $N$ waves.

In a diffractive beam divider, the amplitude and phase field distribution $V(x,y)$ is replaced by a phase-only distribution to maximize the transmitted power. The simplest way to obtain this is, of course, to realize a transparency whose transmission function has unit amplitude and phase given by the argument of $V(x,y)$ hoping that, upon illumination by an orthogonal plane wave, the transparency would produce the requested $N$ plane waves. Unfortunately, generally speaking, this simplest approach does not work because, first, the wave amplitudes turn out not to be in the same ratios as in the $V$ distribution, and, second, the overall light power going into the $N$ desired waves can be modest, much power flowing into spurious diffraction orders generated by the process itself of eliminating the information contained in the amplitude distribution. 

A process of constrained optimization is then realized \cite{RD1,RD2}  in which instead of the field $V(x,y)$ we consider the following, say $U(x,y)$
\begin{equation}
\begin{array}{c}
\displaystyle   U(x,y)=   \sum_{(n, m)\;\in \;B }  \mu_{nm} \, e^{i K (n x+ my)},   
\end{array}
\label{new00s}
\end{equation}
with $\mu_{nm}$ complex quantities, whose phase distribution determines the transmission function $\tau$ of the beam divider
\begin{equation}
\begin{array}{c}
\displaystyle   \tau(x,y)= e^{i\;{\rm arg}\{ U(x, y)  \}}
= \sum_{n, m=-\infty}^\infty   c_{nm} \; e^{i K (n x+ my)},
\end{array}
\label{new00u}
\end{equation}
where arg stands for argument.
The parameters $\mu_{nm}$ are to be determined through an optimization process \cite{RD1,RD2} in such a way that the new Fourier coefficients
\begin{equation}
\begin{array}{c}
 c_{nm}
 =
 \displaystyle \frac{1}{P^2}
 \int_0^P\int_0^P 
 \tau(x,y)
 \; 
 e^{-i K (nx + my)} 
\, {\rm d}x \, {\rm d}y,
\\
\\
\displaystyle (-\infty<n<\infty,\;-\infty<m<\infty),
\end{array}
\label{new11}
\end{equation}
maximize the quantity  
\begin{equation}
\begin{array}{c}
\displaystyle   \eta =    \sum_{(n, m)\;\in \;B } |c_{nm}|^2 ,
\end{array}
\label{new12}
\end{equation}
under the constraint that
\begin{equation}
\begin{array}{c}
\displaystyle     |c_{n_1m_1}|^2=|c_{n_2m_2}|^2 , \;\,\; [\forall (n_1m_1,\; n_2m_2)\in B]
.
\\
\end{array}
\label{new13}
\end{equation}
More generally, the same technique can be used if the $|c_{nm}|$ are required to be in prescribed ratios \cite{Borghi,Borghi2,RD1,RD2,RD3}.  
Notice that even if we are interested in the coefficients whose index pairs belong to $B${,} infinitely many other coefficients different from zero exist because the transparency (\ref{new00u}) is of phase only. In fact, the only phase element producing a finite number of orders is the one corresponding to a pure prism, i.e., a $N$-plicator with $N=1$ \cite{Gori2,RD1,RD2}.

By Parseval theorem \cite{Pap} we have
\begin{equation}
\begin{array}{c}
\displaystyle  \sum_{n=-\infty}^{\infty} \sum_{m=-\infty}^{\infty}   |c_{nm}|^2= 1.
\\
\end{array}
\label{new133}
\end{equation}
The $\eta$ parameter defined in Eq. (\ref{new12}) then represents the energy fraction that flows into the $N$ desired orders. As such, it gives the efficiency of the divider. Common values are around 80\%, but values exceeding 90 \% have been reached \cite{Gori1,Gori2,Borghi,Borghi2,RD1,RD2}.

Note that, since the field distribution $U(x,y)$ of varying amplitude and phase has to be transformed into a pure phase object, the more limited the amplitude variations are in $U$, the greater the efficiency is expected to be \cite{Gori2,RD1,RD2}. This goal is pursued by acting on the $\mu_{nm}$ parameters.

The $N$ pairs $[(n,m) \in B]$  can be taken as coordinates of $N$ points in a suitable plane (physically they could correspond to spots in the far zone) and can be thought of as denoting $N$ distinct propagation directions. A specific choice of the $N$ pairs then corresponds to a particular pattern of directions (or spots).

%\vspace{.3cm}

A significant example, which is of current interest \cite{Moreno19,Moreno22,Moreno23,Vega21,eye,BOE}, is the triplicator. This was the first divider treated with techniques of the calculus of variations \cite{Gori1}. 
In its original version, the three spots were aligned along a unique axis (1D triplicator). Using the present formalism, we take the three spots as aligned on the horizontal axis in the Fourier plane, $K/2\pi$ apart from one another, so that they correspond to the following $(n,m)$ pairs: $(0,0)$, $(1,0)$, $(-1,0)$. Since the problem is one-dimensional, the quantities appearing in all previous equations can be considered as functions of $x$ only, and the parameters can be specified by only one index (namely $c_n$ and $\mu_n$).

The field giving rise to the optimum triplicator is obtained starting from a field that, up to a proportionality factor, is of the form
\begin{equation}
\begin{array}{rl}
U_{\rm 1D}(x)
&
= 
1 + \mu \, e^{i \, K x} + \mu \, e^{-i \, K x}
\\ 
&
=
1 + 2 \mu \, \cos(K x)
\; ,
\end{array}
\label{trip1} 
\end{equation}
where $\mu_{0}$ has been set to one and, for symmetry reasons, we let $\mu_{1}=\mu_{-1}=\mu$. 
The value of $\mu$ that maximizes the efficiency of the element under the uniformity {constraint} (Eq.~(\ref{new13})) has been shown to be $\mu=1.3286 \, i$ \cite{Gori1,Moreno19}, for which $|c_0|=|c_{1}|=|c_{-1}|\simeq 0.555$, corresponding to an efficiency of about 92.6\%.

The same approach can be used to optimize a 2D triplicator, which gives rise to three non-aligned spots in the Fourier plane. As an example of a non-degenerate 2D case, we take the pairs $(0,1)$, $(1,0)$, and $(-1,0)$, so that, on applying the same considerations done in the previous case (i.e., $\mu_{01}=1$, $\mu_{10}=\mu_{-10}=\mu$), we write the field $U$ as
\begin{equation}
\begin{array}{rl}
U_{\rm 2D}(x,y)
&
=
e^{i K y} + \mu \, e^{i \, K x} + \mu \, e^{-i \, K x}
\\
&
=  e^{i K y}  + 2 \mu \cos (K x)
\; .
\end{array}
\label{trip3} 
\end{equation}
The optimization process in this case leads to a maximum efficiency of about 82.6 \% (with $|c_{01}|=|c_{10}|=|c_{-10}|\simeq 0.525$), obtained for any choice of $\mu$, provided that $|\mu|=1$. Therefore, passing from the 1D to the 2D case, the maximum efficiency is considerably reduced. 

An intuitive hint about a possible origin of the discrepancy between the 1D and 2D case can be gained as follows \cite{Gori2,RD2}. As we said, when we replace amplitude and phase field distributions with pure phase functions, the most efficient results can be expected to appear when the original object has the least irradiance variations. 

For the 1D triplicator, the irradiance associated with the field $U_{\rm 1D}$ is
\begin{equation}
\begin{array}{c}
\displaystyle I_{\rm1D}(x)= 1 + a^2 \cos^2(K x),
\end{array}
\label{trip2} 
\end{equation}
where we put $2 \mu = i a$, with $a=2 \times 1.3286= 2.6572$, and varies 
{from 1 to the approximate value of 8.}
%between 1 and 8. 
It never vanishes because the two addends in Eq.~(\ref{trip1}) are in quadrature. Instead, in the 2D case we take $U_{\rm 2D}$ from Eq.~(\ref{trip3}) with $\mu=i$ (for { an easier} comparison with the 1D case) and obtain
\begin{equation}
\displaystyle I_{\rm 2D}(x,y)= 1+ 4 \cos^2 (K x)  + 4\cos (K x)  \sin (K y) 
\; ,
\label{trip4z} 
\end{equation}
which varies between 0 and 9 across the $xy$ plane, over a range which is larger than that of the previous case. Moreover, it is interesting to note that now the irradiance vanishes at some points (e.g., when $Kx=\pi/3$ and $Ky=-\pi/2$), while the one of the field exiting the optical element cannot. Of course, the same behavior of the 1D irradiance is found along the line $y=0$ (where the two addends in Eq.~(\ref{trip3}) are in quadrature) but if $K \ne 0$ you always find points across the plane where the irradiance vanishes. 

%\vspace{.3cm}

\begin{figure}[!ht]
\centering
{\includegraphics[width=8cm]{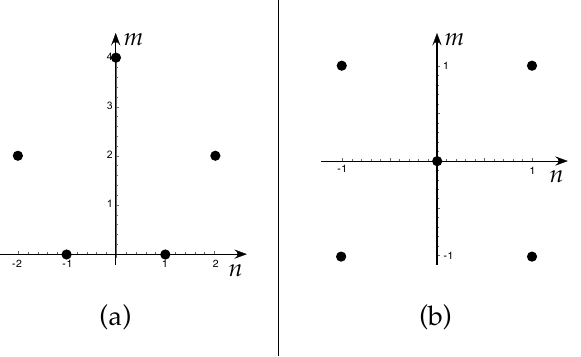}}
\caption{Two different sets of five points on the Fourier plane. Optimization of the $\mu_{nm}$ parameters leads to DOEs with a maximum efficiency of 81.7\% (a) and  85.2\% (b). }
\label{figura2}
\end{figure}

A remarkable and inspiring property of 2D triplicators emerges from their numerical optimization for different shapes of the triangle their diffraction orders describe in the Fourier plane. {In fact, using the same optimization technique that led to Eq.~(\ref{trip3}) it turns out that the same value of the maximum efficiency ($82.6 \%$) is obtained wherever the vertices of the triangle are, as long as they are not aligned.} {That this occurs will be seen to be a consequence of a general rule, the Affinity Rule, that will be exposed in the next Section}.
The same { invariance} property is not exhibited, in general, by beam dividers with $N>3$. 
For example, if we consider the case $N=5$ and arrange the five points as in Fig.~\ref{figura2}(a), after the optimization procedure, the efficiency turns out to be about 81. 7\% (obtained with $\mu_{-1,0}=\mu_{1,0}=e^{i \pi/4}$; $\mu_{2,2}=\mu_{-2,2}=0.9667$; $\mu_{0,4}=1.0578\; e^{i\pi/4}$).
Such an array has a pentagonal shape (even if not regular), and as far as we know, was not studied before. 

We can compare this result with that obtained for the five-point array depicted in Fig.~\ref{figura2}(b), which was studied by Romero and Dickey~\cite{RD2} getting an efficiency of about $84.3\%$. 
In fact, a slightly higher efficiency ($85.2\%$) is reached if the value $1.30405 \; e^{i \pi/4}$ is used for $\mu_{0,0}$ (i.e., for the central point), and the sequence $1, i, i, 1$ for the points {(1,1), (-1,1), (1,-1), (-1,-1)}, respectively.
The fact that, different from the triplicator case, the maximum efficiencies of the two arrays of Fig.~\ref{figura2} are different shows that such a quantity generally depends on the array shape for fixed $N$. 

We can also compare 1D and 2D efficiencies for five-point arrays. For this we cite that the efficiency found in Ref.~\cite{RD1} for five points equally spaced along a line was about $92.1\%$. We then see that the 1D and 2D efficiencies do not coincide, the same as it happened in the case of the triplicator.

%=====================
\section{The affinity rule}
\label{affinity}
%======================

\begin{figure}[h!]
\centering
{\includegraphics[width=8cm]{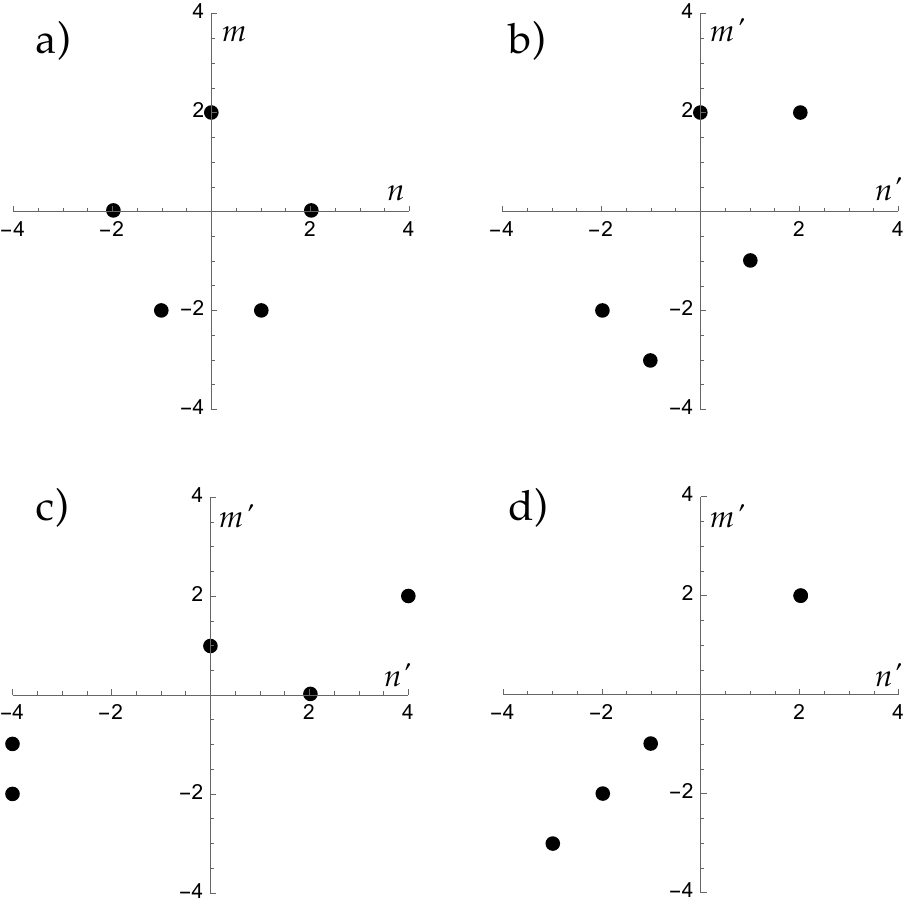}}
\caption{Different sets of five points (b,c,d) obtained by applying the transformations in Eq.~(\ref{trip4}) to the set in (a).}
\label{figura5}
\end{figure}

\begin{figure}[h!]
\centering
{\includegraphics[width=8cm]{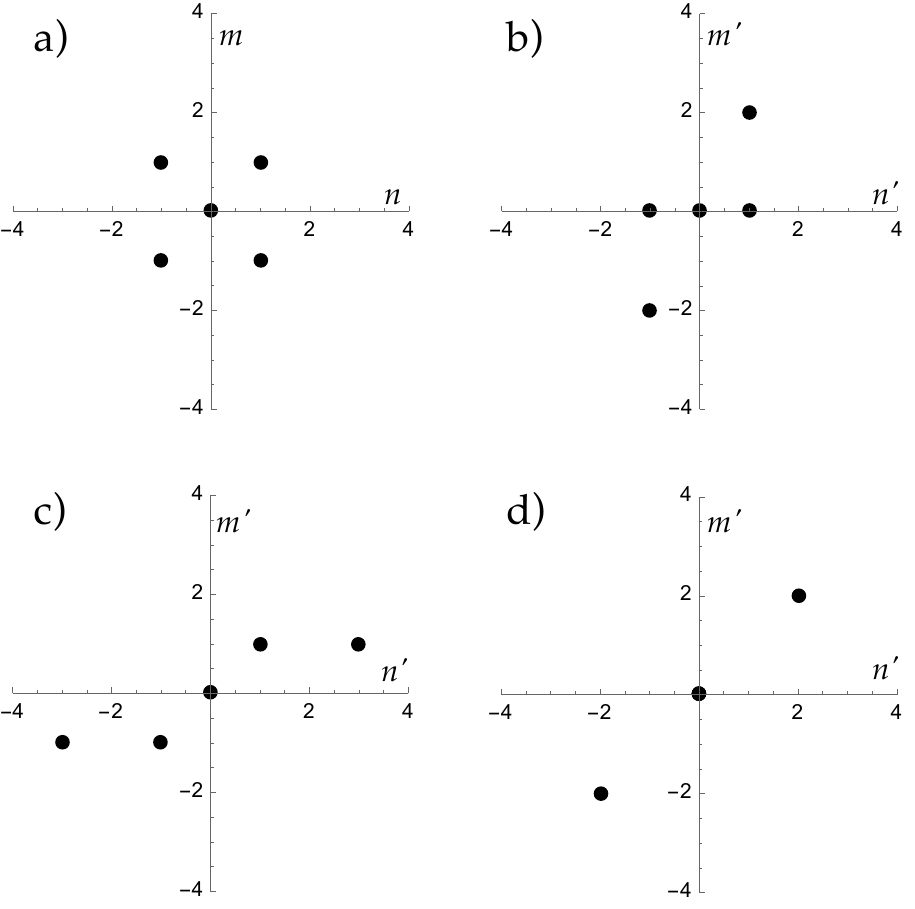}}
\caption{Different sets of five points (b,c,d) obtained by applying the transformations in Eq.~(\ref{trip4}) to the set in (a).}
\label{figura6}
\end{figure}

We can now introduce a rule by which the above results can be interpreted.
Let us suppose we have optimized a two-dimensional $N$-plicator and that a transmission function of the form in Eq.~(\ref{new00u}) has been obtained, with a certain set of $c_{nm}$ coefficients.
{Let us examine the variations that occur in the DOE when an affinity is applied.
We remind that an affinity~\cite{Polanco,Bracewell} is a geometric transformation that preserves lines and parallelism. Consequently, the number of lines that connect points is conserved. Specific examples will be seen in Figs.~\ref{figura5} and \ref{figura6}, where affine and nonaffine transformations will be illustrated.}
Since pure translations can be taken into account by a suitable choice of the reference frame, we limit ourselves to considering the ones with fixed origin, leading from the $xy$ coordinate system to a new one $x'y'$ as follows:
\begin{equation}
\left\{
\begin{array}{l}
x' = a_{11} \, x + a_{12} \, y \; ,
\\
y'=  a_{21} \, x + a_{22} \, y \; ,
\end{array}
\right.
\label{DOE2}
\end{equation}
with real $a_{jk}$ ($j, k = 1, 2$). The equation system is assumed to be invertible ($a_{11}a_{22} - a_{12}a_{21}\ne 0$).

The inverse relations are
\begin{equation}
\left\{
\begin{array}{l}
x = A_{11} \, x' + A_{12} \, y' \; ,
\\
y=  A_{21} \, x' + A_{22} \, y' \; ,
\end{array}
\right.
\label{DOE3}
\end{equation}
where $A_{jk}$ ($j, k = 1, 2$) are the elements of the inverse of the matrix $\{a_{jk}\}$.

When Eqs.~(\ref{DOE3}) are used in Eq.~(\ref{new00u}), the latter gives
\begin{equation}
\begin{array}{rl}
\tau(x, y)
=
&
\displaystyle \sum
c_{nm}
\;
e^{{\rm i}K[n(A_{11} \, x' + A_{12} \, y')+m(A_{21} \, x' + A_{22} \, y')]}
\\ \\
=
&
\displaystyle \sum
c'_{n'm'}
\;
e^{{\rm i}K(n' x' + m' y')}
\\ \\
 =
 &
 \tau'(x', y')
 \; ,
 \end{array}
 \label{DOE4}
\end{equation}
with 
\begin{equation}
\left\{
\begin{array}{l}
n' = A_{11} \, n + A_{21} \, m \; ,
\\
m'=  A_{12} \, n + A_{22} \, m \; ,
\end{array}
\right.
\label{DOE5}
\end{equation}
and $c'_{n'm'} = c_{nm}$. This is consistent with the general rule for Fourier coefficients under affine transformations~\cite{Bracewell}.

{We have supposed $n'_i$ and $m'_i$ ($i=1,2$) to be integers, so that they can be interpreted as the indices of a Fourier series. Of course, this is not guaranteed by a general transformation. However, if the values $n'_i$ and $m'_i$ provided by Eq.~(\ref{DOE5}) are (or can be approximated by) rational numbers, it is always possible to multiply the whole transformation matrix by a scalar quantity and turn them into integers with the same ratio. This corresponds to performing a scaling of the transverse plane.} 

In conclusion, Eq.~(\ref{DOE4}) can be read as follows. Starting from any given periodic DOE, specified by its transmission function $\tau$ (i.e., by a set of Fourier coefficients $c_{nm}$, optimized according to the optical function the DOE has to perform), an infinite number of different DOEs can be conceived, having transmission function $\tau'$, characterized by the same Fourier coefficients as $\tau$, but differently distributed across the Fourier plane. Physically, this means that, if we spatially deform the grating across the plane $xy$ according to an affine transformation, the propagation direction of the diffracted orders changes, but their amplitudes do not. As a consequence, the efficiency of the two gratings is the same. This is what we call the \emph{affinity rule}.

%\vspace{0.3cm}

Let us give some examples, starting from the triplicator. We want to design a DOE that produces three identical replicas of an incident beam, the replicas being directed along three given directions. {We let one of these directions to coincide with} the $z$ axis, the most general case being obtainable by using a simple prism, or adding an equivalent phase to the phase of the DOE. The two remaining directions will be specified by the corresponding spatial frequencies, expressed in terms of the inverse of the grating period. 

The affinity rule allows us to use the results obtained in the optimization of a certain 2D triplicator and extend them to any other 2D triplicator. This can be done starting, for example, from the optimized triplicator (with transmission function $\tau$) that produces three replicas with frequencies $P_0=(0,0)$, $P_1=(n_1,m_1)$, and $P_2=(n_2,m_2)$, and looking for the transmission function ($\tau'$) that produces the three replicas in $P'_0=(0,0)$, $P'_1=(n'_1,m'_1)$, and $P'_2=(n'_2,m'_2)$. 
{As mentioned above, if $n'_i$ and $m'_i$ ($i=1,2$) are not integers but can be approximated by rational numbers, we can always find integer values of the point coordinates such that the shape of the triangle they form is preserved, although its area has changed. In an experimental setup, a scaling factor in the Fourier plane can be compensated for, for example, by using two lenses in a telescopic arrangement.}

%We suppose $n'_i$ and $m'_i$ ($i=1,2$) to be integers, but the results can be easily extended to rational numbers. In such a case, in fact, we can always find integer values of the point coordinates such that the shape of the triangle they form is preserved although its area has changed. A scaling factor in the Fourier plane could be compensated in an experimental setup, for example, by using two lenses in a telescopic arrangement.

First of all, we look for the linear transformation that takes the points $P_1, P_2$ into the points $P'_1, P'_2$ (recall that $P_0=P'_0$). To this aim, we introduce the matrix
\begin{equation}
\widehat A
=
\left(
\begin{array}{cc}
A_{11} & A_{12}
\\
A_{21} & A_{22}
\end{array}
\right)
\; ,
\end{equation}
and the two matrices
\begin{equation}
\widehat M
=
\left(
\begin{array}{cc}
n_{1} & n_{2}
\\
m_{1} & m_{2}
\end{array}
\right)
,
\;\;\;
\widehat{M}'
=
\left(
\begin{array}{cc}
n'_{1} & n'_{2}
\\
m'_{1} & m'_{2}
\end{array}
\right)
\; .
\end{equation}

In such a way,  Eq.~(\ref{DOE5}) can be written as
\begin{equation}
\widehat{M}'
=
\widehat{A}^\intercal
\,
\widehat{M}
\; ,
\end{equation}
so that
\begin{equation}
\widehat{A}^\intercal
=
\widehat{M}'
\,
\widehat{M}^{-1}
\; 
\end{equation}
($\intercal$ denoting the transposed), which allows us to evaluate the matrix $\widehat{A}$.
Of course, the matrix $\widehat{M}$ is required to be invertible, so that the starting points cannot be aligned with the origin. Since the matrix $\widehat{A}$ is required to be non-singular as well, also the determinant of $\widehat{M}'$ cannot be zero. 

In conclusion, for any given pair of points, any other pair of points can be reached by means of an affine transformation, provided that both pairs of points are not aligned with the origin. The latter is a fixed point of the transformation and represents the third vertex of a triangle on the Fourier plane. 
This agrees with a general theorem concerning affine transformations, according to which any two triangles can be changed to one another by means of a suitable affine transform~\cite{Polanco}. 

Then, according to the affinity rule, all nondegenerate triangles have the same efficiency. A similar result holds for parallelograms in the sense that all of them are equivalent up to an affine transformation, be they orthogonal or not, because affinity preserves parallelism. 
Other quadrangles where an equivalence can be established are trapezoids, whose characterizing property is the parallelism of a pair of sides only. 
One property of an affine transformation is that the ratio of lengths of the parallel line segments is invariant. This means that we cannot map a given trapezoid to any other, but the map works only if the bases' ratio is preserved, which limits us to some subset of affine-generated trapezoids.

Once the matrix $\widehat{A}$ is known, the transmission function of the new DOE, obtained from Eq.~(\ref{DOE4}), is a stretched version of the first one. Its Fourier coefficients remain the same, but their frequencies move across the Fourier plane according to the rule in Eq.~(\ref{DOE5}), and the diffraction efficiency is preserved.

\begin{figure}[h!]
\centering
{\includegraphics[width=8cm]{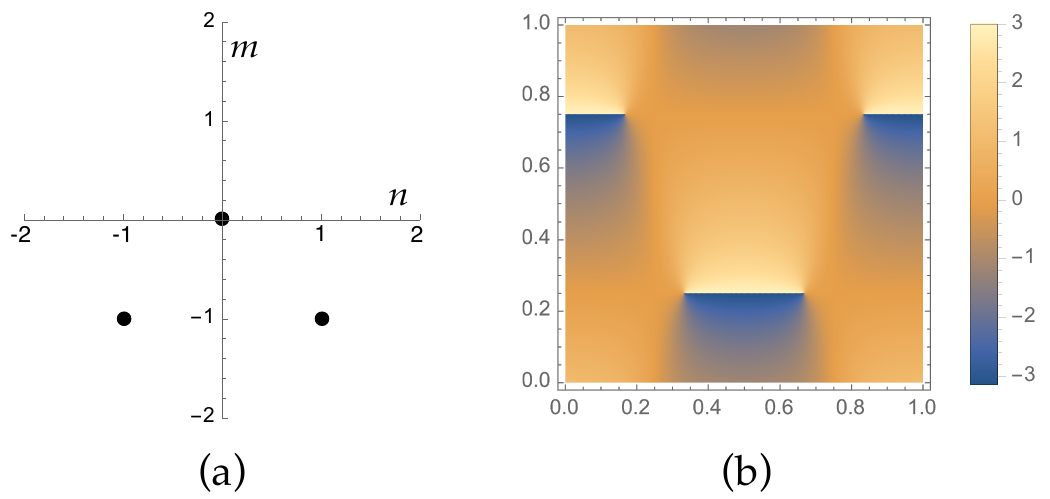}}
\caption{The three diffraction orders produced by a triplicator (a) and the phase profile that maximizes its efficiency (b). {The phase in a unit cell is shown as a function of $(2 \pi x/K,2 \pi y/K)$.}}
\label{figura3}
\end{figure}

Returning to the triplicator, we consider the one involved in Eq.~(\ref{trip3}), which gave rise to an efficiency of $82.6\%$. We take $\mu=i$ and, to have one of the points at the axes origin, we multiply the field $U_{\rm 2D}$ by $\exp(- i K y)$, so that the starting points are at $P_0=(0,0)$, $P_1=(-1,-1)$, and $P_2=(1,-1)$ (see Fig.~\ref{figura3}a). The phase profile of the corresponding optimum triplicator, say $\Phi(x,y)$, is
\begin{equation}
{\Phi(x,y)
=
{\rm Arg}
\left[
1 + 2  i  \cos (K x) \; e^{- i K y} 
\right]}
\, ,
\end{equation}
which is shown in Fig.~\ref{figura3}b as a density plot.

\begin{figure}[h!]
\centering
{\includegraphics[width=8cm]{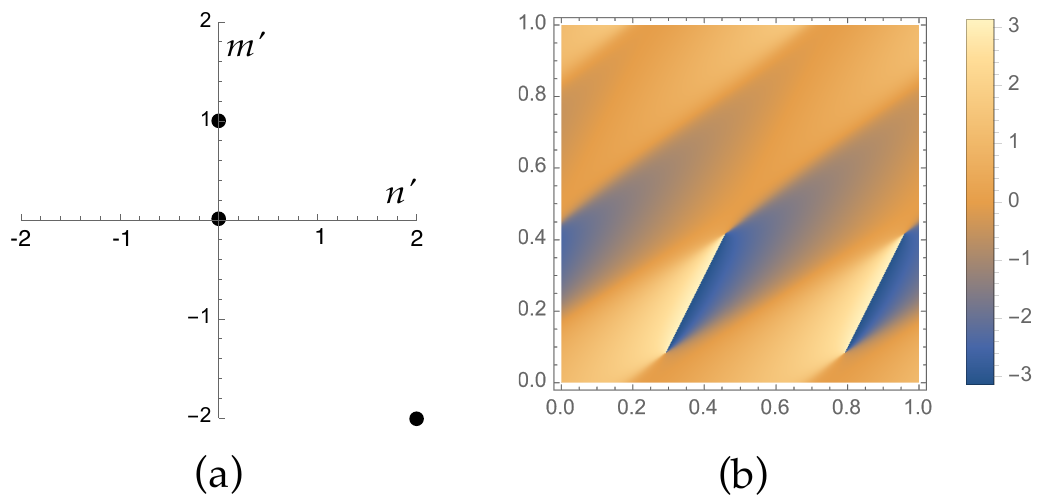}}
\caption{{ Diffraction orders (a) produced applying the affinity rule to the triplicator of Fig.~\ref{figura3} and the corresponding optimum phase profile within a unit cell (b).}}
\label{figura4}
\end{figure}

Now we move the three points to the positions $P'_0=(0,0)$, $P'_1=(0,1)$, and $P'_2=(2,-2)$ (Fig.~\ref{figura4}a). 

The matrices $\widehat M$ and $\widehat M'$ are
\begin{equation}
\widehat M
=
\left(
\begin{array}{rr}
-1 & 1
\\
-1 & -1
\end{array}
\right)
,
\;\;\;
\widehat{M}'
=
\left(
\begin{array}{rr}
0 & 2
\\
1 & -2
\end{array}
\right)
,
\end{equation}
so that 
\begin{equation}
\widehat A
=
\left(
\begin{array}{rr}
1 & -3/2
\\
-1 & 1/2
\end{array}
\right)
\; .
\end{equation}
According to Eqs.~(\ref{DOE4}) and (\ref{DOE3}), the phase profile turns out to be the one shown in Fig.~\ref{figura4}b. The Fourier coefficients can be evaluated numerically, giving rise to the same efficiency as that of the first DOE. 

%\vspace{.3cm}

When $N$ exceeds 3, it is not always possible to move a given set of points in the Fourier plane onto another given set of points. We mentioned some cases where it is possible (for example, when $N=4$ for points at the vertices of a parallelogram), but they are exceptions. What happens, in general, is that for any starting $N$-ple, a whole class of new $N$-ple can be found, all of them being connectable by an affine transform. Elements belonging to the same class can be named \emph{affine} DOEs. The maximum diffraction efficiency for affine DOEs is the same. 
Therefore, it appears sufficient to optimize the phase profile of a single element of the affine DOE class while obtaining other elements by affine transformations.

Let us consider again the pentagonal pattern of {Fig.~\ref{figura2}(a)}.  
If such an array is acted on by an (arbitrary) affine transform described by one of the following matrices:
\begin{equation}
\widehat A_b =
\left(\begin{array}{cc}
\displaystyle  1 & 1\\
\\
\displaystyle 0 & 1
\end{array}
\right)
,
\;\;\;
\widehat A_c =
\left(\begin{array}{cc}
\displaystyle  2 & 1\\
\\
\displaystyle 1 & 0
\end{array}
\right)
,
\;\;\;
\widehat A_d =
\left(\begin{array}{cc}
\displaystyle  1 & 1\\
\\
\displaystyle 1 & 1
\end{array}
\right)
,
\;\;\;
\label{trip4} 
\end{equation}
the new arrays  shown in  Fig.~\ref{figura5} are obtained [a) original; $j$) $\widehat A_j$ ($j=b,c,d$)].
The matrix $\widehat A_d$ is singular and cannot be associated with an affine transformation. 
In fact, two of the starting points (2,0) and (0,2) merge to the same point (2,2), while the rest of the points align.
As it can be seen, the shape change is considerable even if, except for the degenerate case, the new arrays will be recognized as a deformed versions of the array of Fig.~\ref{figura5}a. The efficiency remains unchanged for patterns $a$, $b$, and $c$. The affinity rule does not apply to the pattern $d$, where the transformation brings all points on one line.

Elements from another class of affine DOEs are shown in Fig.~\ref{figura6}. Now the starting pattern is the one already encountered in Fig.~\ref{figura2}(b). The transformation matrices are the ones in Eq.~(\ref{trip4}), and the same considerations hold as for the previous case. In this case also, the transformation $d$ brings all the points onto a single line (and three of them, namely, (0,0), (1,-1), and (-1,1) into the same point, (0,0)).
%

%=====================
\section{Conclusions}
\label{conclusions}
%======================

One of the difficulties encountered in designing beam dividers with diffractive phase transparencies is that, due to the nonlinear character of reducing an amplitude and phase transparency to a phase distribution only, it is generally necessary to make recourse to numerical procedures. Since the efficiency of the device is desired to be as high as possible, maximization processes are to be implemented. This is time consuming and often {ingenuity-driven procedure to be repeated} for any choice of the $N$ directions. Even changing only one of the $N$ directions generally requires repetition of the entire maximization procedure.

We have presented an affinity rule, according to which, once a given divider has been designed, other significant variants can be derived at no cost, i.~e., without repeating the whole optimization process. In such cases, the new DOEs are obtained with the same efficiency, simply by applying a coordinate transformation across the grating plane. We have also shown that transformations that do not correspond to an affinity give rise to new DOE's with different efficiency values.

We limited ourselves to transmission periodic functions in rectangular coordinates. A promising extension could be N-plicators in polar form,giving rise, for example, to diffraction orders at the vertices of a regular polygon in the Fourier plane~\cite{szabo}. Because of their symmetry, this is expected to represent an important class of DOEs for practical applications.

{We note on finishing that our results rely on conditions of scalar diffraction theory. However, there are instances when electromagnetic effects can become significant, for example, when the affinity transform leads to a significant distortion in the $x-y$-plane so that the periods or the diffracting structure becomes rather small compared to the wavelength. Then the values of diffraction efficiencies may become different from those based on the scalar theory.}

%=====================
\section*{Funding}
%=====================

The work has been supported by Spanish Ministerio de Econom\'ia y Competitividad, project PID2019-104268 GB-C21.

%=====================
\section*{Disclosures}
%=====================

The authors declare no conflicts of interest.

%	\begin{backmatter}
%		\bmsection{Funding} Spanish Ministerio de Econom\'ia y Competitividad, project PID2019-104268 GB-C21. OK acknowledges the Copper Fellowship program at the University of Miami.
%	
%		\bmsection{Disclosures} The authors declare no conflicts of interest.
%		
\section*{Data availability} No data were generated or analyzed in the presented research.
%		
%		%\bmsection{Supplemental document} See Supplement 1 for supporting content. 
%		
%	\end{backmatter}


\begin{thebibliography}{99}


\bibitem{Loewen}
E. C. Loewen and E. Popov,
{\em Diffraction Gratings and Applications} 
(CRC Press, Boca Raton, 1997)
%
\bibitem{Soifer}
V.A. Soifer ed.,
{\em Computer Design of Diffractive Optics} 
(Woodhead Publishing, Oxford, 2013)
%
%{
\bibitem{RD3} 
L. A. Romero and Fred M. Dickey, 
Laser Beam Splitting Gratings, in {\em Laser Beam Shaping} Fred M. Dickey, ed. (CRC Press, 2017). 
%}
%
\bibitem{Damm} 
G. H. Dammann and G\"ortler, 
High efficiency in-line multiple imaging by means of multiple imaging phase holograms, 
Opt. Commun. {\bf 3} (1971) 312-315.
%
\bibitem{Jari} 
J. Turunen, A. Vasara, J. Westhrolm, G. Jin, and A. Salin, 
Optimization and fabrication of beam splitters, 
J. Phys. D {\bf 21} (1986) s202-s205.
%
\bibitem{Wyro}  
F. Wyrowski, 
Upper bound of the diffraction efficiency of diffractive phase elements, 
Opt. Lett. {\bf16} (1991) 1915-1917.
%
\bibitem{Gori1} 
F. Gori, M. Santarsiero, S. Vicalvi, R. Borghi, G. Cincotti, E. Di Fabrizio, M. Gentili, 
Analytical derivation of the optimum triplicator, 
Opt. Commun. {\bf 157} (1998) 13-16.
%
\bibitem{Gori2} 
F. Gori,  
Diffractive optics: an introduction, 
in {\em Diffractive optics and optical microsystems} S. Martellucci and A. N. Chester, eds. (Plenum, 1997).
%
\bibitem{Borghi}
R. Borghi, G. Cincotti, and M. Santarsiero, 
Diffractive variable beam-splitter: optimal design, 
J. Opt. Soc. Am. A, {\bf 17} (2000) 63-67.
%
\bibitem{Borghi2} 
R. Borghi, F. Frezza, L. Pajewski, M. Santarsiero, and G. Schettini, 
Optimum even-phase four-beam multiplier, 
Optical Engin. {\bf 41} (2002) 2736-2742.
%
\bibitem{RD1} 
L. A. Romero and Fred M. Dickey,  
Theory of optimal beam splitting by phase gratings. I. One-dimensional gratings, 
J. Opt. Soc. Am. A, {\bf 24} (2007) 2280-2295.
%
\bibitem{RD2}  
L. A. Romero and Fred M. Dickey,  
Theory of optimal beam splitting by phase gratings. II. Square and hexagonal gratings, 
J. Opt. Soc. Am. A, {\bf 24} (2007) 2296-2312.
%
\bibitem{Capasso} s
N. A. Rubin, Z. Shi, and F. Capasso, 
Polarization in diffractive optics and metasurfaces,  
Adv. Opt. Phot.    {\bf 13}   (2021) 836-970. 
%
\bibitem{Moreno19} 
D. Marco, M. M. S\'anchez-L\'opez, A. Cofr\'e, A. Vargas, and I. Moreno, 
Optical triplicator design applied to a geometric phase vortex grating, 
Op. Ex. {\bf 27} (2019) 14472-14486.
%
\bibitem{Moreno22} 
E. Nabadda, P. Garc\'ia-Mart\'inez, M. M Sa\'nchez-L\'opez, and I. Moreno, 
Phase-Shifting Common-Path Polarization Self-Interferometry for Evaluating the Reconstruction of Holograms Displayed on a Phase-Only Display, 
Front. Phys. {\bf 10} (2022) 920111.
%
\bibitem{Moreno23} 
E. Nabadda,  M. del Mar S\'anchez-L\'opez, P. Garc\'ia-Mart\'inez,  and I. Moreno, 
Retrieving the phase of diffraction orders generated with tailored gratings, 
Opt. Lett. {\bf 48} (2023) 267-270.  
 %
\bibitem{Pap}
A. Papoulis, {\em Systems and transforms with applications in optics}, (McGraw-Hill, 1968). 
% 
\bibitem{Vega21} 
F. Vega, M. Valentino, F. Rigato, and M. S. Mill\'an, 
Optical design and performance of a trifocal sinusoidal diffractive intraocular lens, 
Biomed. Op. Ex. {\bf 12} (2021) 3338-3351.
%
\bibitem{eye} 
F.  Vega, M. Faria-Ribeiro, J. Armengol and M. S. Mill\'an, 
Pitfalls of using NIR-based clinical instruments to test eyes implanted with diffractive intraocular lenses,  
Diagnostic {\bf 13} (2023) 1259-1275.
%
\bibitem{BOE}  
Y. Xing, Y. Liu, K. Li, X. Li, D. Liu, Y. Wang , 
Approach to the design of different types of intraocular lenses based on an improved sinusoidal profile, 
Biomed Op. Ex. {\bf 14}  (2023)  2821-2838.
%
\bibitem{Byron} 
F. W. Byron, Jr, R. W. Fuller, {\em Mathematics of classical and quantum physics}, (Dover, 1992).
%
\bibitem{Polanco} 
C. Polanco, {\em Transformations: A Mathematical Approach}, (Bentham, 2018).
%
\bibitem{Bracewell} R. N. Bracewell, K. -Y. Chang, A. K. Jha and Y. -H. Wang, Affine theorem for two-dimensional Fourier Transform, Electr. Lett. {\bf 29} (1993) 204.
%
\bibitem{szabo} 
S. Szabo, 
Affine regular polygons,
Elem. Math. {\bf 60} 137-147 (2005).

\end{thebibliography}
\end{document}